\documentclass{ifacconf}

\usepackage{graphicx}      % include this line if your document contains figures
\usepackage{natbib}        % required for bibliography
\usepackage{macros}
\usepackage{algorithm}% http://ctan.org/pkg/algorithm
\usepackage{algpseudocode}% http://ctan.org/pkg/algorithmicx
\usepackage{color}
\usepackage{url}
\begin{document}
\begin{frontmatter}

\title{Applications Oriented Input Design in Time-Domain Through Cyclic Methods\thanksref{footnoteinfo}}
% Title, preferably not more than 10 words.

\thanks[footnoteinfo]{The research leading to these results has received funding from the European Union’s Seventh Framework Programme (FP7/2007-2013) under grant agreement no 257059, the ’Autoprofit’ project (www.fp7-autoprofit.eu). This work was partially supported by the Swedish Research Council under contract 621-2009-4017.}
\author[First]{A. Ebadat}
\author[First]{B. Wahlberg}
\author[First]{H. Hjalmarsson}
\author[First]{C. R. Rojas}
\author[First]{P. H\"{a}gg}
\author[First]{C. A. Larsson}

\address[First]{ACCESS Linnaeus Center, School of Electrical Engineering, KTH – Royal Institute of Technology, SE-100 44 Stockholm, Sweden (e-mail: \{ebadat, bo, hjalmars, crro, pehagg, chrisla\}@kth.se).}
%\address[Second]{Colorado State University,
%   Fort Collins, CO 80523 USA (e-mail: author@lamar. colostate.edu)}
%\address[Third]{Electrical Engineering Department,
%   Seoul National University, Seoul, Korea, (e-mail: author@snu.ac.kr)}

\begin{abstract}                % Abstract of not more than 250 words.
In this paper we propose a method for applications oriented input design for linear systems under time-domain constraints on the amplitude of input and output signals. The method guarantees a desired control performance for the estimated model in minimum time, by imposing some lower bound on the information matrix. The problem is formulated as a time-domain optimization problem, which is non-convex. This is addressed through an alternating method, where we separate the problem into two steps and at each step we optimize the cost function with respect to one of two variables. We alternate between these two steps until convergence. A time recursive input design algorithm is performed, which enables us to use the algorithm with control. Therefore, a receding horizon framework is used to solve each optimization problem. Finally, we illustrate the method with two numerical examples which show the good ability of the proposed approach in generating an optimal input signal.
\end{abstract}


\end{frontmatter}
%===============================================================================

\section{Introduction}
System identification concerns the problem of data-based plant modeling and plays an important role in industry. One of the key enabling issues in any system identification problem is the choice of input signal. An appropriate input signal should be able to extract as much useful information as possible from the system. Therefore, a properly designed input signal can improve the quality of the identified model, significantly. This problem has led to formation of the topic \emph{optimal input design}. 

Optimal input design has been extensively investigated in the literature, see e.g. \cite{L.Ljung1999}, \cite{Goodwin&Payne:77}, and \cite{Bombois&Gilson:06SYSID}. The problem has been formulated in many different forms, however, one common idea is to design an input signal such that a certain accuracy is obtained during the identification while the experimental effort to obtain such an accuracy is minimized. This accuracy is often defined in terms of the application of the model and thus the identification objective is to guarantee that the estimated model belongs to the set of models that satisfies the desired control specifications, with a given probability. This induces growth of the ideas of identification for control, least costly identification and applications oriented input design, see \cite{hjalmars:04}, \cite{Gevers&Ljung:86}, \cite{Bombois&Gilson:06SYSID}, and \cite{Hjalmarsson2009}. The problem is usually defined as an optimization problem where one tries to satisfy the requirements on the quality of the model by using the minimum experimental effort. 
The quality of the model can be measured by the Fisher matrix, which determines the amount of information regarding the model and, the inverse of this matrix is a lower bound on the covariance matrix for any unbiased estimator (\cite{L.Ljung1999}). 
\\
For model structures linear in input, the information matrix is asymptotically an affine function of the input power spectrum. Therefore, the input design problem is usually formulated in the frequency domain and the outcome is an optimal input spectrum or an autocorrelation sequence. The optimal input values are obtained from the given optimal spectrum, see \cite{Fedorov:72}. The problem is, however, more complex for nonlinear dynamical systems, since the input spectrum is not enough to describe the information matrix. Here, the probability density function of the input signals can be optimized instead of the input spectrum. Similar to the linear case, one can generate the time realizations given the probability density function, see e.g., \cite{Pato2013}, and \cite{Forgione2014}.

In practice there are some constant bounds on the input signals and the resulting output signals, which should be taken into account during the experiment design. These constraints are typically expressed in the time-domain and how to handle this in frequency-domain is not evident. One way to get around this problem is to impose these constraints during the generation of a time realization of the desired input spectrum, see \cite{Gujar:68}, \cite{Liu:82}, \cite{Schroede1970} and \cite{LarssonPer2013} .

There are, however, some approaches that try to solve the optimal input design problem in the time domain directly, see e.g \cite{Manchester} for linear systems. The main advantage is that in the time domain the constraints on the amplitude of the input and the system dynamics appears naturally and are easier to handle. However, the main difficulty that arises is that the problem is non-convex. In \cite{Manchester}, this problem is addressed through a semidefinite relaxation of quadratic programs and the Fisher information matrix is maximized under some constraints on the input signal. 

In this paper we propose a novel method for applications oriented input design for linear systems in the time domain, where it is straightforward to formulate the constraints on both the input and output signals. The aim is to satisfy some lower bound on the information matrix  in minimum time, which can guarantee a desired control performance for the estimated model. The problem is formulated as an optimization problem by adding a positive slack variable to the lower bound. The problem, however, is non-convex, which imposes a high computational burden. We try to get around the non-convexity through alternating methods. More precisely, we solve the problem for one variable, when the other is fixed. Thus, in this paper we separate the problem into two steps where at each step we optimize the cost function with respect to one set of the variables and we alternate between these two steps until convergence. In order to detect the minimum required time the problem is formulated in a receding horizon manner and we perform a time recursive algorithm. This enables us to overcome the computational burden and makes the formulation suitable to be used with controllers such as Model Predictive Control (MPC). The input signal is optimized over a prediction horizon, but only the first value is applied to the plant and the optimization is performed again in the next step. This is repeated until the lower bound on the information matrix is satisfied.

The outline of the paper is as follows. In Section \ref{Sec:Problem Formulation}, we go through the formulation of the problem and the necessary mathematical background. We describe the optimal input design problem in Section \ref{Sec:Optimal Input Design}, followed by a detailed description of the proposed new method in Section \ref{sec:Time-Domain Optimal Input Design}.  Section \ref{FIR example} gives a better insight into the proposed algorithm by describing the method for Finite Impulse Response models. In Section \ref{sec:Numerical Results}, we illustrate the method in two numerical examples and in Section \ref{sec:conclusion}, some conclusions are stated.

\subsection{Notation}
 $\mathbb{E}\{.\}$ denotes the expected value. We use $( . ) ^{\frac{1}{2}}$ to denote a Hermitian square root of a positive definite matrix. Define $\mathcal{S}^+_n$ to be the set of positive $n \times n$ semi-definite matrices. A matrix $A \in \mathbb{R}^{m \times n}$ with $m>n$, is said to be semi-unitary if $A^T A = I_{n \times n}$, where $I_{n \times n}$ is an $n \times n$ identity matrix.
%===================================================================================================================================
\section{Problem Formulation}
\label{Sec:Problem Formulation}
Consider the identification of discrete-time multivariate systems that are causal linear time-invariant (LTI)
\begin{equation} \label{Eq:System}
				y(t) = G_0(q)u(t) + H_0(q)e_0(t),
\end{equation}
where $u(t) \in \R^{n_u}$ and $y(t)\in\R^{n_y}$ are the input and output vectors and $e_0(t)\in\R^{n_e}$ is white Gaussian noise with zero mean and covariance matrix $\Lambda$. $G_0(q)$ and $H_0(q)$ are the transfer function matrices of the system. Let $q^{-1}$ denote the backward shift operator, e.g., $q^{-1}u(t) = u(t-1)$.
\\
In system identification, we want to find a model of the system \eqref{Eq:System}. We assume that the model is parametrized by an unknown parameter vector $\theta \in \R^{n_\theta}$, that is,
\begin{equation}  \label{Eq:Model}
	\mathcal{M}(\theta): \quad		y(t) = G(q,\theta)u(t) + H(q,\theta)e(t).
\end{equation}
In addition, we assume that the model \eqref{Eq:Model} matches system \eqref{Eq:System} exactly when $\theta=\thetao$. We call $\thetao$ the true parameter vector. The objective of system identification is to estimate the value of $\theta$ that best describes the system, according to some quality measure. The estimated parameter vector, given $N$ measurements in the experiment, is denoted $\hat{\theta}_N$.

It is assumed that the estimated model will be used by a model-based controller like Model Predictive Controller (MPC) where the model is employed to find the predicted output used in the MPC cost function. The more accurate the model, the better the controller performance will be. In the next section we will describe the main idea behind optimal input design which is considered in this paper.
%
%===================================================================================================================================
\section{Optimal Input Design}
\label{Sec:Optimal Input Design}
The idea in optimal input design in the least costly framework is to minimize an experimental effort, such as input power, while satisfying some requirements on the accuracy of the identified model. These requirements can be expressed in terms of control performance. The problem is then to design an input signal to be used in the identification experiment such that the estimated model guarantees acceptable control performance when used in the control design, namely \emph{applications oriented experiment design} (see \cite{Bombois&Gilson:06SYSID} and references therein). 
\subsection{Application Cost}
We use the concept of application cost function to relate the plant-model mismatch to the performance degradation. We use a scalar function of $\theta$ as the application cost and denote it $\Vapp(\theta)$.
The cost function is chosen such that its minimum value occurs at $\theta = \thetao$. In particular, we assume without loss of generality that $\Vapp(\thetao) = 0$. Note that if $\Vapp(\theta)$ is twice differentiable in a neighborhood of $\thetao$, this implies that
%\begin{center}
	$\Vapp(\thetao) = 0$ , $\Vapp'(\thetao) = 0$ and $\Vapp''(\thetao) \geq 0.$
%\end{center}
%
There are many possible choices of application functions with these properties, see e.g. \cite{Larsson2011a}.
\\
A maximum allowed performance degradation imposes an upper bound on the application cost function, that is
\begin{equation} \label{Eq:ApplicationCost}
	\Vapp(\theta) \leq \frac{1}{\gamma},
\end{equation}
where $\gamma$ is a user-defined positive constant. Each parameter vector $\theta$ that fulfills the inequality \eqref{Eq:ApplicationCost}, can be considered as an acceptable parameter from an application's point of view. Therefore, the set of all acceptable parameters, namely the \emph{application set}, is defined as
\begin{equation}\label{Eq:ApplicationSet}
	\Theta (\gamma) = \left\{ \theta : \Vapp(\theta) \leq \frac{1}{\gamma} \right\}.
\end{equation}
We can make a local convex approximation of $\Theta(\gamma)$ by invoking the Taylor expansion of $\Vapp(\theta)$ around $\thetao$ and considering its mentioned properties around $\thetao$:
\begin{equation} \label{Eq:ellipsoidalApprox}
\begin{split}
	\Vapp(\theta) &\approx
									 \Vapp(\thetao) + \Vapp'(\thetao) [\theta - \thetao]\\
						& +  0.5 [\theta - \thetao]^T\Vapp''(\thetao) [\theta - \thetao]\\
						&= 0 + 0 + 0.5 [\theta - \thetao]^T \Vapp''(\thetao) [\theta - \thetao].
\end{split}
\end{equation}
Thus we have the following ellipsoidal approximation of the application set (see~\cite{Hjalmarsson2009}):
\begin{equation}\label{Eq:ellipsoidalApproxApplicationSet}
	\Theta (\gamma) \hspace{-0.1cm} \approx  \hspace{-0.1cm} \Eapp(\gamma) \hspace{-0.1cm}=\hspace{-0.1cm}  \left\{\theta: [\theta - \thetao]^T V^{''}_{app}(\thetao) [\theta - \thetao]  \leq \frac{2}{\gamma} \right\}.
\end{equation}
\vspace{-0.3cm}
%~~~~~~~~~~~~~~~~~~~~~~~~~~~~~~~~~~~~~~~~~~~~~~~~~~~~~~~~~~~~~~~~~~~~~~~~~~~~~~~~~~~~~~~~~~~~~~~~~~~~~~~~~~~~~~~~~~~~~~~~~~~~~~~~~~
\subsection{System Identification}
We use the prediction error method (PEM) with quadratic cost to estimate the unknown parameters of the considered system, $\theta\in\R^n$, from $N$ available samples of input-output data, see \cite{L.Ljung1999}. A key asymptotic ($N\rightarrow \infty$) property of PEM, is that the estimated parameters lie in an \emph{identification set} with a certain probability say $\alpha,$ (\cite{Wahlberg&Ljung:92}). This set is defined as
\begin{equation}
	\label {id set}
	\Esi(\alpha)=\left\{
							\theta:[\theta-\thetao]^TI_F(\thetao)[\theta-\thetao]\leq  {\chi^2_\alpha(n_\theta)}
					 \right\},
\end{equation}
where $\chi^2_\alpha(n)$ is the $\alpha$-percentile of the $\chi^2$-distribution with $n$ degrees of freedom and $I_F$ is the Fisher information matrix, which measures the amount of information regarding the unknown parameters, $\theta$, that can be obtained from observations of the output signal. We thus have that $\hat{\theta}_N\in\Esi(\alpha)$ with probability $\alpha$ when $N\rightarrow \infty$. In this paper we assume that $N$ is finite but sufficiently large such that asymptotic properties hold. For more details, we refer the reader to \cite{L.Ljung1999}.
%~~~~~~~~~~~~~~~~~~~~~~~~~~~~~~~~~~~~~~~~~~~~~~~~~~~~~~~~~~~~~~~~~~~~~~~~~~~~~~~~~~~~~~~~~~~~~~~~~~~~~~~~~~~~~~~~~~~~~~~~~~~~~~~~~~
\subsection{Applications Oriented Experiment Design}
As mentioned before, in applications oriented input design, the input signal used in the identification experiment is designed such that the estimated model guarantees acceptable control performance when used in the control design, that is, it requires that $\hat{\theta}_N \in \Theta(\gamma)$  with high probability. One way to ensure this is to require
\begin{align} \label{Eq:ExperimentDesign}
	\Esi(\alpha) \subseteq  \Theta(\gamma).
\end{align}
Using this set constraint, the input design problem can be formulated as an optimization problem, where \eqref{Eq:ExperimentDesign} plays the role of a constraint. In order to make the problem convex, the ellipsoidal approximation of the application set, \eqref{Eq:ellipsoidalApproxApplicationSet}, can be used in \eqref{Eq:ExperimentDesign}. Thus, both sets are ellipsoids and the problem becomes the following linear matrix inequality (LMI) in the elements of $I_F$:
\begin{equation} \label{Eq:LMI}
	\frac{1}{\chi^2_\alpha(n_\theta)}I_F(\thetao) \geq \frac{\gamma}{2}V^{''}_{app}(\thetao).
\end{equation}
Finally, a natural objective in the input design is to minimize an experiment cost, such as input power or energy or experimental time, while \eqref{Eq:LMI} is fulfilled, i.e.
\vspace{-0.4cm}
\begin{eqnarray} \label{Eq:OID}
\begin{split}
&\min_{\substack {\text{input}}} \quad 	
								&&\text{Experimental Cost}\\
&\text{s.t.} \quad
								&&\frac{1}{\chi^2_\alpha(n_\theta)}I_F(\thetao) \geq \frac{\gamma}{2}V^{''}_{app}(\thetao).
\end{split}
\end{eqnarray}
\\
Since $I_F$ is an affine function of the input spectrum in open loop identification (\cite{L.Ljung1999}), the constraint \eqref{Eq:LMI} can be formulated as \emph{LMIs} by linear parameterization of the input spectrum. Therefore, the optimization problem \eqref{Eq:OID} is usually solved for the input spectrum and thus the outcome of the optimization is often not given as a sequence of values but rather it is given as an optimal input spectrum or an autocorrelation sequence. 
%%For a moving average stationary signal of order $M$, the spectral density can be written as:
%%
%%\begin{equation} \label{Eq: OptimalInputSpectrum}
%%\Phi_u(\omega) = \sum_{k=-M}^{k=M}c_k\beta_k(e^{j\omega}),
%%\end{equation}
%%
%%where $\{\beta_k(e^{j\omega})\}$ are basis functions. One choice of basis functions is $\beta_k(e^{j\omega})=e^{-j\omega}$, corresponding to
%%
%%\begin{equation} \label{Eq:coefficients for OID}
%%c_k = E\{u(t)u^T(t-k) \}.
%%\end{equation}
%%

Notice that, in practice usually the maximum allowed amplitudes for inputs and outputs values are restricted by the process. These constraints on inputs and outputs stem from the saturation of the actuator and the need to keep the system within a safe operation region, respectively. It is of great significance that these constraints are also being satisfied during the system identification experiment.
\rem{Note that we do not have the knowledge of the true parameters, $\thetao$. However, this can be addressed by either implementing a robust experiment design scheme on top of it (see \cite{Rojas2007}) or through an adaptive procedure where the calculations of the Hessian of the cost function and output predictions are updated as more information is being collected from the system, see e.g. \cite{Gerencser2009}, and \cite{Per2013}. In the sections to follow, we will use the true parameter vector. In practice this has to be replaced by an initial estimate.}

%===================================================================================================================================
\section{Time-Domain Optimal Input Design}
\label{sec:Time-Domain Optimal Input Design}
In this paper we introduce a new solution to the applications oriented input design. The objective is to satisfy the constraint \eqref{Eq:LMI} in \emph{minimum time} while we are forcing the input and output signals to lie in certain convex sets. In the context of the problem \eqref{Eq:OID}, the experimental cost here is the minimum required time to satisfy the experiment constraint. The problem is formulated as a time-domain optimization problem, where it is straightforward to handle time-domain constraints on input and output signals by solving the problem directly in the time-domain. 

To be able to formulate the problem, we define a new slack variable. The constraint \eqref{Eq:LMI} is then satisfied if there exists a positive semidefinite matrix $S$ such that: 
\begin{equation} \label{Eq:S}
I_F(\thetao) - \frac{\chi^2_\alpha(n_\theta)\gamma}{2}V^{''}_{app}(\thetao) - S = 0, \quad S\geq 0,
\end{equation}
where $S$ is a positive semi-definite slack variable. We then try to minimize $J = {\left \| I_F(\thetao) - \frac{\chi^2_\alpha(n_\theta)\gamma}{2}V^{''}_{app}(\thetao) - S   \right \|}_F^2$ for $S\geq 0$, under input and output constraints. Here, $\|.\|_F$ denotes the Frobenius norm. The experiment design constraint is satisfied if we obtain $J = 0$.

In order to find the minimum required time, we perform a time recursive input design algorithm. This also makes the algorithm compatible to be used with controllers such as MPC since they are using the same context. Hence, we formulate the input design problem as the following receding horizon problem, where at each time $t$, we solve
\vspace{-0.4cm}
\begin{eqnarray} \label{Eq:OID Time-Domain}
\begin{split}
&\min_{\substack {\{ u(k)\}_{k=t}^{t+N_u},S}}
								&& J_t = {\left \| I_F^{t+N_u}(\thetao) - \frac{\chi^2_\alpha(n_\theta)\gamma}{2}V^{''}_{app}(\thetao) - S  \right \|}_F^2\\
&\text{s.t.}
			  			&& S \geq 0 ,\\
			      		&&& u(k) \in \mathcal{U}, \quad k = t, \ldots, t+N_u,\\
						&&& y(k) \in \mathcal{Y}, \quad k = t, \ldots, t+N_y,\\
						&&& y(k) = G(q,\theta_0)u(k)  , \quad k = t, \ldots, t+N_y.
\end{split}
\end{eqnarray}
Here, $\mathcal{U}$ and $\mathcal{Y}$ are convex constraint sets on the input and the output, respectively. These could for example correspond to system amplitude constraints. $N_u$ and $N_y$ are input and output horizons. If $N_y$ is longer than $N_u$ the input is considered zero over the rest of the output horizon. In this paper, we assume $N_u = N_y$.$I_F^{t+N_u}(\thetao) $ is the Fisher information matrix up to time $t+N_u$, see Section \ref{Sec:Fisher Information Matrix}. Although the solution to the problem \eqref{Eq:OID Time-Domain} is a sequence of input values, we only apply the first value to the system and the optimization is performed again in the next time step, according to the receding horizon principle.

At each time sample $t$, if the lower bound on the information matrix is fulfilled, i.e. $I_F^{t+N_u}(\thetao) \geq \frac{\chi^2_\alpha(n_\theta)\gamma}{2}V^{''}_{app}(\thetao)$, then $J_t = 0$ holds and vice versa.  We can then stop running the receding horizon \eqref{Eq:OID Time-Domain} when $J_t = 0$ holds for the first time and consider this time to be the minimum time required to satisfy the application requirements.

To iteratively solve \eqref{Eq:OID Time-Domain}, we first need to rewrite the information matrix $ I_F^{t+N_u}(\thetao)$ in a recursive form and relate it to the input $u(t)$. Then a cyclic algorithm is proposed to address the input design problem \eqref{Eq:OID Time-Domain}.
%\rem {The problem could also be formulated as the dual problem of \eqref{Eq:OID}. This is achieved by making the sets $\mathcal{U}$ and $\mathcal{Y}$ gradually smaller until there exists no input sequence that leads to $J_t = 0$ in finite time. As a result, we will get the minimum experimental cost with which we can satisfy the experiment design constraint. This requires adding an outer loop to the problem \eqref{Eq:OID Time-Domain} and solving \eqref{Eq:OID Time-Domain} for different sets $\mathcal{U}$ and $\mathcal{Y}$ at each iteration of the loop.}
\rem{The formulation \eqref{Eq:OID Time-Domain} can also be used to find the maximum accuracy $\gamma$, for which we can satisfy \eqref{Eq:LMI} in the sets $\mathcal{U}$ and $\mathcal{Y}$. To this end, one can solve \eqref{Eq:LMI} for different values of $\gamma$ and increase $\gamma$ until there exists no input sequence that leads to $J_t = 0$ in finite time. This requires adding an outer loop to \eqref{Eq:OID Time-Domain} and solving it for different values of $\gamma$ at each iteration of the loop.}
 \vspace{-0.05cm}
%~~~~~~~~~~~~~~~~~~~~~~~~~~~~~~~~~~~~~~~~~~~~~~~~~~~~~~~~~~~~~~~~~~~~~~~~~~~~~~~~~~~~~~~~~~~~~~~~~~~~~~~~~~~~~~~~~~~~~~~~~~~~~~~~~~
\subsection{Fisher Information Matrix}
\label{Sec:Fisher Information Matrix}
For an unbiased estimator, the inverse of the Fisher matrix is a lower bound on the covariance of the parameter estimation error, according to Cram\'{e}r-Rao bound. The information matrix is (\cite{Goodwin&Payne:77}):
\begin{equation}
I_F(\theta) := \mathbb{E} \Big \{
						\frac{\partial \log p(y|\theta)}{\partial \theta}  \frac{\partial \log p(y|\theta)}{\partial \theta}^T
						\Big\} \ \in \R^{n_{\theta} \times n_{\theta} }.
\end{equation}
Considering the model \eqref{Eq:Model} and assuming $e(t)$ to be a Gaussian white noise, the log likelihood function is:
\vspace{-0.0cm}
\begin{equation}
\log p(y|\theta) = \text{constant} - \frac{1}{2} \sum_{t =1}^{N} \epsilon^T(t,\theta) \Lambda^{-1}\epsilon(t,\theta)
\end{equation}
\vspace{-0.0cm}
where $N$ is the number of samples that are being used in the computation of the Information matrix and $\epsilon(t,\theta) \in \R^{n_y}$ is the prediction error given by
\begin{equation}
\nonumber
\epsilon(t,\theta) := H^{-1}(q,\theta) [ y(t) - G(q,\theta) u(t) ].
\end{equation}
Assume that the plant and noise models are parameterized independently and let $\theta_G \in \R^{n_{\theta_G}}$ denote the parameters of the system model while $\theta_H \in \R^{{n_\theta}_H}$ contains the parameters of the noise model. The Fisher information matrix for data up to time $t+N_u$ is
\begin{equation}
{I}_F^{t+N_u}(\theta)  \hspace{-0.1cm} := \hspace{-0.2cm}  \sum_{k=1}^{k = t+N_u}  \hspace{-0.2cm}  \mathbb{E} \Bigg \{\hspace{-0.1cm}
																				\begin{bmatrix}
																					\frac{\partial \epsilon^T(t,\theta)}{\partial \theta_G}\\
																					\\
																					\frac{\partial \epsilon^T(t,\theta)}{\partial \theta_H}\\
																				\end{bmatrix} \hspace{-0.1cm}
																						\Lambda^{-1}
																			\begin{bmatrix}\hspace{-0.2cm}
																					\frac{\partial \epsilon^T(t,\theta)}{\partial \theta_G}\\
																					\\
																					\frac{\partial \epsilon^T(t,\theta)}{\partial \theta_H}\\
																				\end{bmatrix}^T  \hspace{-0.2cm} \Bigg \}
\end{equation}
where, $\frac{\partial \epsilon^T(t,\theta)}{\partial \theta_G} \in \R^{n_{\theta_G} \times n_y}$ and $\frac{\partial \epsilon^T(t,\theta)}{\partial \theta_H} \in \R^{{n_\theta}_H \times n_y}$. Now if we assume that $\left\{u(t)\right\}$ and $\left\{e(t)\right\}$ are uncorrelated (i.e. the system is operating in open loop), we obtain
\vspace{-0.2cm}
\begin{eqnarray}
{I}_F^{t+N_u}(\theta) = \mathbb{E}  \{
						\begin{bmatrix}
							\bar{I}_F^{t+N_u}(\theta_G) & 0\\
								0 &  \bar{I}_F^{t+N_u}(\theta_H)
%							\frac{\partial \epsilon^T(t,\theta)}{\partial \theta_G} \Lambda^{-1} \frac{\partial \epsilon(t,\theta_G)}{\partial \theta_G}  & 0\\
%							0 & \frac{\partial \epsilon^T(t,\theta)}{\partial \theta_H}\Lambda^{-1}\frac{\partial \epsilon^T(t,\theta)}{\partial \theta_H}
						\end{bmatrix} \}.
\end{eqnarray}
Since $I_F(\theta_H)$ only depends on the noise $e(t)$, the only part of information matrix that can be optimized by the choice of input signal is
\vspace{-0.05cm}
\begin{equation}
\bar{I}_F^{t+N_u}(\theta_G) =  \hspace{-0.3cm} \sum_{k=1}^{k = t+N_u}
																			\left( \frac{\partial \epsilon^T(t,\theta)}{\partial \theta_G} \right)
																			\Lambda^{-1}
																			\left( \frac{\partial \epsilon^T(t,\theta)}{\partial \theta_G} \right)^T,
\end{equation}
considering that $\mathbb{E} \{\bar{I}_F^{t+N_u}(\theta_G)\} = \bar{I}_F^{t+N_u}(\theta_G)$, since $\frac{\partial \epsilon^T(t,\theta)}{\partial \theta_G}$ is deterministic. On the other hand, one can write
\vspace{-0.0cm}
\begin{eqnarray} \label{SplitFIM}
\begin{split}
\bar{I}_F^{t+N_u}(\theta_G) &:= \sum_{k=1}^{k = t-1}
																			\left( \frac{\partial \epsilon^T(t,\theta)}{\partial \theta_G} \right)
																			\Lambda^{-1}
																			\left( \frac{\partial \epsilon^T(t,\theta)}{\partial \theta_G} \right)^T
										\\&+
											\sum_{k=t}^{k = t+N_u}
																			\left( \frac{\partial \epsilon^T(t,\theta)}{\partial \theta_G} \right)
																			\Lambda^{-1}
																			\left( \frac{\partial \epsilon^T(t,\theta)}{\partial \theta_G} \right)^T.
\end{split}
\end{eqnarray}
\vspace{-0.0cm}
The first term in \eqref{SplitFIM} depends on the values of the input signal up to time $t-1$, which are assumed to be known at time $t$. Therefore, we focus on the second term, which contains the inputs in the horizon in the optimization problem \eqref{Eq:OID Time-Domain}.
 Based on the definition of $\epsilon(t,\theta)$
\begin{equation}
\frac{\partial \epsilon^T(t,\theta)}{\partial \theta_G} =
																	\begin{bmatrix}
																		\mathcal{F}_1(q)u(t)\\
																		\mathcal{F}_2(q)u(t)\\
																		\vdots \\
																		\mathcal{F}_{n_{\theta_G}}(q)u(t)\\
																	\end{bmatrix}
																		,
\end{equation}
where $n_{\theta_G}$ is the number of parameters in the model and
\begin{equation}
\mathcal{F}_i (q)u(t) = - \Big [ H^{-1}(q, \theta_H) \frac{\partial G(q, \theta_G)}{\partial (\theta_G(i))}u(t) \Big]^T.
\end{equation}
\\
Building on \cite{Manchester}, the elements of the reduced information matrix can be written as:
\begin{eqnarray} \label{Eq:FIMelements}
\begin{split}
(\bar{I}_F^{t+N_u})_{i,j}(\theta_G) &=\hspace{-0.1cm} (\bar{I}_F^{t-1})_{i,j}(\theta_G)\\
												&	+	\hspace{-0.1cm}\sum_{k=t}^{k = t+N_u} (\mathcal{F}_{i} (q)u(k))  \Lambda^{-1}  (\mathcal{F}_{j}(q) u(k)),
\end{split}
\end{eqnarray}
where $i,j = 1, \ldots, n_{\theta_G}$ and $ (\bar{I}_F^{t-1})_{i,j}(\theta_G)$ is obtained using available data at time instant $t$.
Denote the impulse response of $\mathcal{F}_i(q)$ by $f_i(t)$, the maximum length of the truncated impulse responses of $\mathcal{F}_i(q)$ for $i = 1, \ldots, n_{\theta_G}$ by $n$, and define $(\tilde{I}_F^{t-1})_{i,j}(\theta_G)$ as the part of the information matrix depending on the future values of $u(t)$. Define
\begin{equation} \label{Eq:F_i}
F_i :=
		\begin{bmatrix}
			f_i(n)  &  f_i(n-1) &   \ldots    &   f_i(1)     			    			&   \ldots 		& 0 \\
				0    &   f_i(n)  &    \ldots    &   f_i(2)     			    			&   \ldots 		& 0 \\
			\vdots &  \vdots  &    \ddots   &   \vdots   			   			&   \ddots 	   & \vdots\\
				0    &    0       &   \ldots    &   f_i(N_u)     		   			&   \ldots 		   & f_i(1) \\
		\end{bmatrix},
\end{equation}
%
%
%&   0
%& f_i(1)
%& \vdots
%& f_i(N_u-1)
where $F_i \in \R^{  (N_u+1)n_y  \times  (N_u+n-1)n_u  }$, and
\vspace{-0.2cm}
\begin{eqnarray}
\label{InputVector}
\begin{split}
\bar{u}^*(t) &:= [u^*(t-n+1), \ldots,u^*(t-1)] &&\in \R^{(n-1)n_u},\\
\bar{u}(t)     &:= [u(t), \ldots, u(t+N_u) ] &&\in \R^{(N_u)n_u}.
\end{split}
\end{eqnarray}
The former is already known at time $t$ while we are going to optimize the latter. We can then rewrite \eqref{Eq:FIMelements} as
\vspace{-0.2cm}
\begin{eqnarray}
\begin{split}
(\tilde{I}_F^{t+N_u})_{i,j}(\theta_G) &= \hspace{-0.1cm} \begin{bmatrix}
													(\bar{u}^*(t))^T & \bar{u}(t)^T
													\end{bmatrix} F_i^T \Lambda_e^{-1}F_j
													\begin{bmatrix}
													\bar{u}^*(t) \\
													\bar{u}(t)
													\end{bmatrix}\hspace{-0.1cm},\\
\Lambda_e^{-1} &= I_{(N_u+1)\times(N_u+1)} \otimes \Lambda^{-1},
\end{split}
\end{eqnarray}
see \cite{Manchester}. Eventually we have
\begin{eqnarray}
\begin{split}
\tilde{I}_F^{t+N_u}\hspace{-0.05cm}(\theta_G) \hspace{-0.10cm}	&=\hspace{-0.1cm}	
															\begin{bmatrix}
															\textbf{u}^T F_1^T \Lambda_e^{-1} F_1 \textbf{u} 					& \hspace{-0.10cm}\ldots 	& \textbf{u}^T F_1^T \Lambda_e^{-1} F_{n_{\theta_G}} \textbf{u} \\
																						\vdots                                       							&\hspace{-0.10cm} \ddots 	&                                    \vdots 												 \\
															\textbf{u}^T F_{n_{\theta_G}}^T \Lambda_e^{-1} F_1 \textbf{u} 	& \hspace{-0.10cm}\ldots 	& \textbf{u}^T F_{n_{\theta_G}}^T \Lambda_e^{-1} F_{n_{\theta_G}} \textbf{u}
																\end{bmatrix}
																\hspace{-0.15cm},
\end{split}
\end{eqnarray}
where 
\begin{align}\label{u}
\textbf{u} =\left[(\bar{u}^*(t))^T \ \bar{u}(t)^T \right]^T.
\end{align}
Therefore,
\begin{equation}
\bar{I}_F^{t+N_u}(\theta_G) = \tilde{I}_F^{t+N_u}(\theta_G) + \bar{I}_F^{t-1}(\theta_G).
\end{equation}
Defining
\begin{equation}
\label{Phiu}
\Phi(\textbf{u}) \hspace{-0.1cm} = \hspace{-0.1cm} [\Lambda_e^{-\frac{1}{2}}F_1 \textbf{u} , \ \ldots, \ \Lambda_e^{-\frac{1}{2}}F_{n_{\theta}} \textbf{u}]\ \in \R^{(N_u+1)n_y\times n_{\theta_G}},
\end{equation}
the Fisher information matrix can be written as:
\begin{equation} \label{Eq:QuadraticFIM}
\bar{I}_F^{t+N_u}(\theta_G) =  \Phi(\textbf{u})^T\Phi(\textbf{u})+ \bar{I}_F^{t-1}(\theta_G).
\end{equation}
Since $\Phi(\textbf{u})$ is linear in $\textbf{u}$, one can see that the information matrix is a quadratic function of the input sequence.
\vspace{0cm}
%~~~~~~~~~~~~~~~~~~~~~~~~~~~~~~~~~~~~~~~~~~~~~~~~~~~~~~~~~~~~~~~~~~~~~~~~~~~~~~~~~~~~~~~~~~~~~~~~~~~~~~~~~~~~~~~~~~~~~~~~~~~~~~~~~~~
\subsection{A Cyclic Algorithm}
For simplicity we assume that the application cost function depends only on the plant model. Thus we can use the reduced information matrix in \eqref{Eq:OID Time-Domain}. Substituting \eqref{Eq:QuadraticFIM} into the cost function in \eqref{Eq:OID Time-Domain} and with some abuse of notation
\begin{equation}
J_t = \left \| \Phi(\textbf{u})^T\Phi(\textbf{u})+\textbf{C}(t-1) - S  \right \|_F^2,
\end{equation}
where
\begin{align}
\textbf{C}(t-1) = \bar{I}_F^{t-1}(\theta_0) - \frac{\chi^2_\alpha(n_\theta)\gamma}{2}V^{''}_{app}(\thetao)
\end{align}
is a known matrix at time $t$ which can be computed using data available at time $t$.
The optimization problem \eqref{Eq:OID Time-Domain} is non-convex and is in general hard to solve. However, the cost function is separable in terms of the variables, which makes it possible to find a solution of the problem through alternating algorithms (see e.g. \cite{Tropp2005} and \cite{stoica2007}). To put it another way, we can break the problem into two smaller problems by considering only one of the variables, $\textbf{u}$ and $S$, at each time. The resulting problems are easier to solve. This motivates us to propose a cyclic algorithm for this problem. The method alternates between optimizing the cost function using one of the variables while the other is fixed. Therefore, two main steps are allocated for the proposed algorithm.
\subsubsection{Step1} \label{Algorithm1}
Assuming $S$ is fixed to its most recent optimal value, $S_{opt}$, we aim to solve the following optimization problem at time instant $t$:
\vspace{-0.3cm}
\begin{eqnarray} \label{Eq:Algorithm1}
\begin{split}
&\min_{\substack {\textbf{u}}}
								&& \left \| \Phi(\textbf{u})^T\Phi(\textbf{u})+\textbf{C}(t-1) - S_{opt}  \right \|_F^2\\
&\text{s.t.}
			      		&& \textbf{u} =\left[(\bar{u}^*(t))^T \ \bar{u}(t)^T \right]^T \in \mathcal{U},\\%\quad k = 1, \ldots, n+N_u,\\
						&&& \textbf{u}(k) = \bar{u}^*(t), \quad k = 1, \ldots, n-1,\\
						&&& y(k) \in \mathcal{Y}, \quad k = t, \ldots, t+N_u,\\
						&&& y(k) = G(q,\theta_0){u}(k)  , \quad k = t, \ldots,t + N_u.
\end{split}
\end{eqnarray}
where $\textbf{u}$ is defined in \eqref{u} and $\textbf{u}(k)$ is the $k^{th}$ element of $\textbf{u}$.
The optimization problem \eqref{Eq:Algorithm1} is still not convex. However, the class of unconstrained signals, $\Phi(\textbf{u})$, for which the cost function is zero, is (\cite{stoica2007})
\begin{align} \label{Eq:HRS}
\Phi(\textbf{u}) = U \big( S_{opt} - \textbf{C}(t-1) \big)^{\frac{1}{2}},
\end{align}
if
\begin{align} \label{Eq: PD}
S_{opt} - \textbf{C}(t-1) \geq 0,
\end{align}
where $U \in \R^{(N_u+1)n_y \times n_{\theta_G}}$ is a semi-unitary matrix. We will later show that the property \eqref{Eq: PD} holds at time instant $t-1$. Hence, the problem \eqref{Eq:Algorithm1} can be relaxed to
\vspace{-0.3cm}
\begin{eqnarray} \label{Eq:OID_Step1_Reduced}
\begin{split}
&\min_{\textbf{u},U}
								&& \left \| \Phi(\textbf{u}) - U \big(S_{opt} - \textbf{C}(t-1) \big) ^{\frac{1}{2}}  \right \|_F^2\\
&\text{s.t.}
			      		&& \textbf{u} =\left[(\bar{u}^*(t))^T \ \bar{u}(t)^T \right]^T \in \mathcal{U},\\%\quad k = 1, \ldots, n+N_u,\\
						&&& \textbf{u}(k) = \bar{u}^*(t), \quad k = 1, \ldots, n-1,\\
						&&& y(k) \in \mathcal{Y}, \quad k = t, \ldots, t+N_u,\\
						&&& y(k) = G(q,\theta_0){u}(k)  , \quad k = t, \ldots,t + N_u,\\
						&&& U^T U = I.
\end{split}
\end{eqnarray}
The cost function is still non-convex. However, this problem, in turn can be broken into two problems by considering only one of the variables and fixing the other one. Since $\Phi(\textbf{u})$ is linear in $\textbf{u}$ we will come up with two convex problems in terms of $\textbf{u}$ and $U$. Therefore, we can again use a cyclic optimization algorithm in order to solve the problem. Here, we will use the minimization algorithm suggested in \cite{stoica2007}. The algorithm is alternating between the following two steps until convergence:
\begin{description}
\item{Step 1.1}: Assuming $U$ is fixed to its most recent optimal value, solve the problem \eqref{Eq:OID_Step1_Reduced} for $\textbf{u}$, which is a constrained quadratic programming problem
\vspace{-0.2cm}
\begin{eqnarray} \label{Eq:OID_Step1.1}
\begin{split}
\textbf{u}_{opt} &=&& \arg \min_{\textbf{u}}
								 \left \|
										 \Phi(\textbf{u}) - U_{opt} \big(S_{opt} - \textbf{C}(t-1) \big) ^{\frac{1}{2}} 
									\right \|_F^2\\
&\text{s.t.}
			      		&& \textbf{u} =\left[(\bar{u}^*(t))^T \ \bar{u}(t)^T \right]^T \in \mathcal{U},\\%\quad k = 1, \ldots, n+N_u,\\
						&&& \textbf{u}(k) = \bar{u}^*(t), \quad k = 1, \ldots, n-1,\\
						&&& y(k) \in \mathcal{Y}, \quad k = t, \ldots, t+N_u,\\
						&&& y(k) = G(q,\theta_0){u}(k)  , \quad k = t, \ldots,t + N_u
\end{split}
\end{eqnarray}
\item{Step 1.2}: Having the optimal input sequence, $\textbf{u}_{opt}(t)$, find optimal $U$ for \eqref{Eq:OID_Step1_Reduced} through Singular Value Decomposition (SVD), i.e:
\vspace{-0.1cm}
\begin{eqnarray} \label{Eq:SVD}
\begin{split}
(S_{opt} - \textbf{C}(t-1) \big) ^{\frac{1}{2}} \Phi(\textbf{u}_{opt})^T &= \bar{U}\Sigma \tilde{U}^T, \
																		U_{opt} = \tilde{U}\bar{U}^T.
\end{split}
\end{eqnarray}
See \cite{stoica2007} for more details.
\end{description}
%
%
%The algorithm can be summarized as follows:
%\begin{algorithm}
%  \caption{Optimize $\textbf{u}$ when $S$ is fixed}\label{Algorithm1}
%  \begin{algorithmic}[1]
%   \State Update $U$ to its most recent value
%   \State Solve \eqref{Eq:OID_Step1.1} and obtain $\textbf{u}_{opt}$
%   \State Find $U_{opt}$ using \eqref{Eq:SVD}
%  \end{algorithmic}
%\end{algorithm}
%
\subsubsection{Step2}\label{Step2}
Having obtained the optimal solution, $\textbf{u}_{opt}(t)$, from the first step, we need to solve
\vspace{-0.2cm}
\begin{eqnarray} \label{Eq:OID_Algorithm2}
\begin{split}
&\min_{\substack {S}}
								&& \left \| \Phi(\textbf{u}_{opt})^T\Phi(\textbf{u}_{opt}) + \textbf{C}(t-1) - S  \right \|_F^2\\
&\text{s.t.}
			      		&& S \geq 0.\\
\end{split}
\end{eqnarray}
An important advantage of the proposed algorithm is that we can find a closed-form solution for this step. The optimal solution of \eqref{Eq:OID_Algorithm2} is the projection of $\Phi(\textbf{u}_{opt})^T\Phi(\textbf{u}_{opt}) + \textbf{C}(t-1)$ onto  $\mathcal{S}^+_{n_\theta}$ (\cite{ProjectionMethodsInConicOptimization2012}). To determine this projection note that since $\Phi(\textbf{u}_{opt})^T\Phi(\textbf{u}_{opt}) + \textbf{C}(t-1)$ is symmetric, we can write
\begin{align}\label{Eq:EigenValueDecomposition}
\Phi(\textbf{u}_{opt})^T\Phi(\textbf{u}_{opt}) \hspace{-0.1cm}+  \hspace{-0.1cm} \textbf{C}(t-1) \hspace{-0.1cm}=\hspace{-0.1cm} V \text{diag} (\lambda_1, \ldots, \lambda_{n_{\theta}}) V^T  \hspace{-0.1cm},
\end{align}
where $\lambda_i$ are the eigenvalues and $V$ is the corresponding orthonormal matrix of eigenvectors. Thus
\begin{align} \label{Eq:Sopt}
S_{opt}= V \text{diag} \big(\max(0,\lambda_1), \ldots, \max(0,\lambda_{n_{\theta}}) \big) V^T.
\end{align}
See~\cite{ProjectionMethodsInConicOptimization2012} for further information.
Note that $S \geq \Phi(\textbf{u}_{opt})^T\Phi(\textbf{u}_{opt}) + \textbf{C}(t-1)$ according to \eqref{Eq:Sopt}, which confirms that the property \eqref{Eq: PD} holds.

%This algorithm can also be summarized as follows:
%\begin{algorithm}
%  \caption{Optimize $S$ when $\textbf{u}$ is fixed}\label{Algorithm2}
%  \begin{algorithmic}[1]
%   \State Update $\textbf{u}$ to its most recent value
%   \State Find a solution for \eqref{Eq:OID_Algorithm2} through \eqref{Eq:EigenValueDecomposition}-\eqref{Eq:Sopt}
%  \end{algorithmic}
%\end{algorithm}
As mentioned before, the proposed alternating method in this paper cycles between Step 1 and Step 2. The resulting problem only involves solving a quadratic optimization problem, an SVD of a matrix with size $n_{\theta}$ and a projection and thus it is fast enough to address large problems. 
\rem{We have no proof of convergence yet, for the proposed method, however, from the numerical simulations good convergence results are obtained see also \cite{stoica2007} and references there in for more examples. We also refer to \cite{Tropp2005} for more details and properties of alternating approaches.}\\
The method is summarized in the table below\footnote{One possibility for stopping criteria is to stop the iterations when the tolerance of changes in the variables is small enough.}. 
\begin{algorithm}
\renewcommand{\thealgorithm}{}
  \caption{Proposed Alternating Method}\label{Proposed Alternating Method}
  \begin{algorithmic}[]
	  \State \emph{Initialization:}
	\State \quad choose $N_u$ and n
	\State \quad $S_{opt}\gets 0$, $U_{opt}\gets U_{init}$, $t\gets  1$ and $J_0 \neq 0$	%\Comment{$U_{init}^T U_{init} = I$}
      \While{$J_{t-1} \neq 0$}
			\State i = 1
      		\While{\{Stopping criteria is not true\}}
				\State \emph{Start Step 1:}
				\State \quad Solve \eqref{Eq:OID_Step1.1}
				\State \quad $\textbf{u}_{opt}^i \gets \textbf{u}_{opt}$
				\State \quad Use $\textbf{u}_{opt}^i$ and \eqref{Eq:SVD} to compute $U_{opt}$
				\State \quad ${U}_{opt}^i \gets U_{opt}$
				\State\emph{Start Step 2:}
				\State \quad Use $\textbf{u}_{opt}^i$, ${U}_{opt}^i$ and \eqref{Eq:EigenValueDecomposition}-\eqref{Eq:Sopt} to obtain $S_{opt}$
				\State \quad ${S}_{opt}^i \gets S_{opt}$
				\State $i \gets i+1$
			\EndWhile
        \State $u^*(t)\gets $First sample of the optimal input signal
        \State Calculate $\bar{I}_F^{t}$, $\textbf{C}(t)$ and $J_t$
        \State $t\gets  t+1$
      \EndWhile\label{euclidendwhile}
      \State \textbf{return} optimal input sequence ${\{u^*(k)\}}_{k=1}^{t+N_u}$
  \end{algorithmic}
\end{algorithm}
\vspace{-0.4cm}
%===================================================================================================================================
\section{FIR example}
\label{FIR example}
To get a better insight into the proposed approach, we study it for a simple Finite Impulse Response (FIR) model
\begin{eqnarray} \label{Eq:FIR_Model}
\begin{split}
&y(t,\theta) = \theta_1 u(t-1) + \theta_2 u(t-2) + e(t),\\
&\mathbb{E}\{e(t)\} = 0 \ , \  \mathbb{E}\{e(t)^2\} = \lambda,
\end{split}
\end{eqnarray}
where $\theta =[\theta_1, \ \theta_2]$. Assume we aim to design an optimal input sequence with minimum length such that the identified model based on the obtained input signal can guarantee a desired control performance when it is being used in a controller. Moreover, we assume that according to some physical restrictions we need
\begin{equation}
|u(t)| \leq u_{max}, \ |y(t)| \leq y_{max}.
\end{equation}
In order to solve the problem we use the following steps.
\subsection{Desired control performance}
One reasonable choice of a desired performance for the controller is the difference between the measured output when the controller is working based on the true parameters, $\thetao$, and when it is working based on the estimated parameters, $\hat \theta$, that is,
\begin{equation}\label{Eq:FIR_Vapp}
\Vapp(\hat \theta) = \frac{1}{N} \sum_{t=1}^{N} {\| y(t,\thetao) - y(t,\hat \theta)  \|}^2,
\end{equation}
over a step response of the system with the controller running.
Since for \eqref{Eq:FIR_Vapp}, we have $\Vapp(\thetao) = 0$ and $\Vapp'(\thetao) = 0$,
%\begin{center}
%\end{center}
we can approximate the set \eqref{Eq:ApplicationSet} by \eqref{Eq:ellipsoidalApproxApplicationSet}. The Hessian matrix can be calculated through either numerical or analytical methods, depending on the type of controller. Now, having defined $\Vapp(\thetao)$, we aim to design an input sequence such that \eqref{Eq:LMI} is fulfilled for a given $\gamma$.

\subsection{Input design}
The signal generation is done through the optimization problem \eqref{Eq:OID Time-Domain}. We first need to find the Fisher information matrix. Considering \eqref{Eq:FIR_Model}, we attain
\begin{eqnarray}
\begin{split}
&\epsilon(t) = y(t) - \theta_1 u(t-1) - \theta_2 u(t-2),\\
&\frac{\partial \epsilon^T(t,\theta)}{\partial \theta} =-
																	\begin{bmatrix}
																		q^{-1} u(t)\\
																		q^{-2} u(t)
																	\end{bmatrix}.
\end{split}
\end{eqnarray}
Assume we are at time instant $t$ and we aim to optimize the input signal in the prediction horizon of length $N_u$, putting $n=3$, we can write \eqref{InputVector} and \eqref{Phiu} as
\begin{eqnarray}
\begin{split}
&\bar{u}^*(t) = [u^*(t-2),u^*(t-1)],\\
&\bar{u}(t) = [u(t),\ldots,u(t+N_u)],\\
&\textbf{u} = [u^*(t-2),u^*(t-1),u(t),\ldots,u(t+N_u)],\\
&\Phi(\textbf{u}) =\frac{1}{\sqrt{\lambda}} [F_1 \textbf{u} ,\ F_2 \textbf{u}],
\end{split}
\end{eqnarray}
%
%\begin{equation} \label{FIR_phi}
%
%\end{equation}
where $F_1$ and $F_2$ are obtained using \eqref{Eq:F_i}. For example choosing $N_u = 4$, we have
\begin{equation}
\begin{split}
&F_1 \hspace{-0.15cm}=\hspace{-0.15cm}
		\begin{bmatrix}
			    0    &    -1      &     0     	&     0        	&   0 		&   0	 &   0 \\
				0    &    0       &    -1       &     0     		&   0     	&   0 	 &   0 \\
				0    &    0       &     0       &    -1   		&   0   	&   0 	 &   0 \\
				0    &    0       &     0       &     0     		&  - 1   	&   0  &   0 \\
				0    &    0       &     0       &     0     		&   0   	&  -1  &   0 \\
		\end{bmatrix}\hspace{-0.15cm},
F_2 \hspace{-0.10cm}=\hspace{-0.15cm}
		\begin{bmatrix}
			    -1   &    0       &     0     	&     0        	&   0 		&   0 &   0 \\
				0    &    -1      &     0       &     0     		&   0     	&   0	&   0 \\
				0    &    0       &    -1       &     0   		&   0   	&   0 &   0\\
				0    &    0       &     0       &     -1     		&   0   	&   0 &   0\\
				0    &    0       &     0       &     0     		&   -1   	&   0 &   0\\
		\end{bmatrix}\hspace{-0.15cm},
\end{split}
\end{equation}
and thus
\begin{eqnarray}
\begin{split}
\tilde{I}_F^{t+N_u}\hspace{0cm}(\theta) \hspace{-0.1cm}	&=\hspace{-0.1cm}	\frac{1}{\lambda}  \hspace{-0.1cm} \sum_{k=t}^{t+N_u}  \hspace{-0.15cm}
																\begin{bmatrix}
																 	u(k-1)u(k-1)	&  	u(k-1)u(k-2) \\
																 	u(k-2)u(k-1) 	&  	u(k-2)u(k-2)
																\end{bmatrix}
																\hspace{-0.15cm}.
\end{split}
\end{eqnarray}
The information matrix for the FIR system is determined by the covariances of input sequences (See \cite{Stoica1982}). We are now ready to find the optimal input signal, $\bar{u}(t)$, using the proposed alternating method.

%~~~~~~~~~~~~~~~~~~~~~~~~~~~~~~~~~~~~~~~~~~~~~~~~~~~~~~~~~~~~~~~~~~~~~~~~~~~~~~~~~
\section{Numerical Results}
\label{sec:Numerical Results}
In this section we implement the suggested method on two examples. The first example is the FIR example explained in Section \ref{FIR example}, while for the second example we consider an output error model with four unknown parameters.
\subsection{Example 1}
\begin{figure}[ht]
      \centering
     \includegraphics[scale = 0.55]{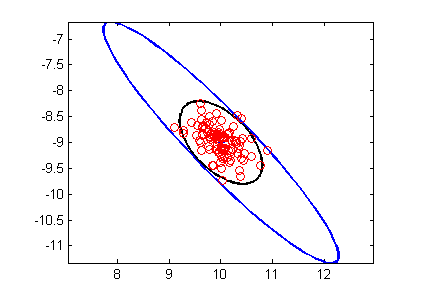}
      \caption{$\Eapp$ is the outer ellipse, $\Esi$ is the inner ellipse and $\hat{\theta}_N$ are the small circles.}
      \label{Figure:EstimatedSamples}
\end{figure}

\begin{figure}[htpb]
      \centering
     \includegraphics[scale = 0.5]{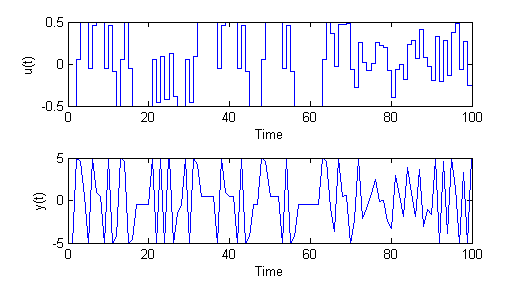}
      \caption{Generated optimal input and output signals. The constraints are satisfied.}
      \label{Figure:OptimalInput}
\end{figure}

\begin{figure}[thpb]
      \centering
     \includegraphics[scale = 0.55]{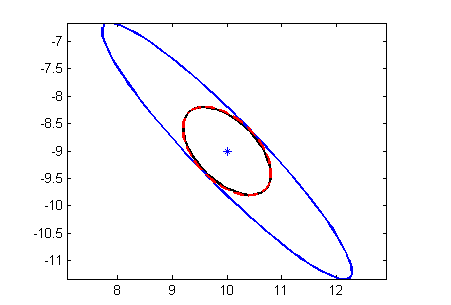}
      \caption{$\Eapp$ is the outer ellipse. The identification ellipse obtained by MOOSE is shown in red ('\textcolor{red}{- -}') and the one obtained by the proposed method in black ('\textcolor{black}{--}').} 
      \label{Figure:Comparison}
\end{figure}

Consider the FIR example in Section \ref{FIR example}, with 
\vspace{-0.4cm}
\begin{eqnarray}
\begin{split}
\nonumber
\thetao 	= [10,\ -9]	, \
%\lambda  =  1 		, \
u_{max} 	= 0.5		, \
y_{max} 	= 5		, \
N_u 		= 5			.
\end{split}
\end{eqnarray}
Assume that we want to generate an input sequence of length $N=100$ that when used in an system identification experiment satisfies both the application requirements and the input and output constraints. The identified model will be used in MPC, with cost function
\begin{equation}
J\hspace{-0.1cm} =\hspace{-0.15cm}\sum^{N_y}_{k=0}\left\|y(k+1)  \hspace{-0.1cm}-\hspace{-0.1cm}  r(k+1)\right\|^{2},
\end{equation}
the same input and output constraints as during the experiment and $r=0$. We calculate the Hessian of the application cost function employing numerical methods, provided by DERIVESTsuite (\cite{DErrico2007}). The required accuracy is $\gamma = 100$ and we want that the estimated parameters lie in the identification set with probability $\alpha = 0.95$. The suggested method is used to obtain an optimal input sequence.  For the obtained input the slack variable $S$, is strictly positive definite, and thus the experiment design constraint is satisfied. The application and identification ellipsoids for the obtained input are shown in Figure \ref{Figure:EstimatedSamples}.
%\begin{figure}[thpb]
%      \centering
%     \includegraphics[scale = 0.6]{Pics/ellipsoids.png}
%      \caption{ $\Eapp$ is the outer ellipsoid and $\Esi$ is the inner ellipsoid. $\Esi$ lies inside $\Eapp$, which means that the estimated parameters will satisfy the application requirements with probability $\alpha$.}
%      \label{Figure:Application_Identification_Sets}
%\end{figure}
The generated input signal has been used in the system identification experiment with zero mean white Gaussian noise $e(t)$ with variance $\lambda = 1$. One hundred $\hat{\theta}_N$ are estimated based on the measurements of $y(t)$, when the obtained input signal is applied to the system. To this aim the system identification toolbox in Matlab is used. In total 95\% of the estimated parameters are inside the identification ellipsoid. The results are shown in Figure \ref{Figure:EstimatedSamples}. It can be seen that $\Esi$ is inside $\Eapp$, thus, the performance requirement will be fulfilled by the estimated parameters with probability 95\%.
\vspace{-0.1cm}
The generated input signal has been shown in Figure \ref{Figure:OptimalInput}. It can be seen that the signal satisfies the constraint. However, it is worth to note that the constraints are only applied on the noiseless output signal. This can be addressed by the same approach discussed in Remark1 and using observers such as Kalman Filter.
\vspace{-0.1cm}
\\
We also formulate the problem in the frequency domain using \eqref{Eq:OID}, where the input power is chosen to be the experimental cost. We use MOOSE, a toolbox for optimal input design implemented in MATLAB (\cite{moose}), in order to solve problem \eqref{Eq:OID}. Result is shown in Figure \ref{Figure:Comparison}. However, as mentioned before what we are getting out of solving \eqref{Eq:OID} is an input spectrum and we need to find the corresponding time realization by another optimization problem which is not an easy problem under input and output constraints (see  \cite{LarssonPer2013}).
\vspace{-0.1cm}
\vspace{-0.2cm}
\subsection{Example 2}
Consider the output error model of a two tank system:
\begin{equation}
\begin{split}
x(t+1) &= 
\begin{bmatrix}
\theta_3 && \theta_4\\
1           &&     0     
\end{bmatrix}
x(t) +
\begin{bmatrix}
4.5 \\
0        
\end{bmatrix}
u(t) ,\\
y(t) &= 
\begin{bmatrix}
\theta_1 &&  \theta_2
\end{bmatrix}
x(t) + e(t).
\end{split}
\end{equation}
The upper tank is connected to a pump with input $u(t)$. The tank has a hole in the bottom with free flow into a lower tank, which also has a hole with free flow out of the tank. The level in the lower tank is the output, $y(t)$. The true system parameters are given by $[0.12 \ 0.059 \ 0.74 \ -0.14]^T$. Again assume that we aim to generate an input signal with length N=100 such that the identified model satisfies the application requirements. Other parameters are 
%\begin{align*}
$\lambda  =  0.01 		, \
u_{max} 	= 0.5		, \
y_{max} 	= 5		,$ and
$N_u 		= 5			$.
%\end{align*}
The application cost function is assumed to be \eqref{Eq:FIR_Vapp}, where the controller is MPC with the following cost function
\begin{equation}J\hspace{-0.1cm} =
															\hspace{-0.15cm}\sum^{N_y}_{k=0}\left\|y(k+1)  \hspace{-0.1cm}-\hspace{-0.1cm}  r(k+1)\right\|^{2}_{Q}\hspace{-0.1cm}
															+\hspace{-0.1cm}\sum^{N_u}_{k=1}\left\|\Delta u(k)\right\|^{2}_{R} \hspace{-0.3cm},
\end{equation}
where, $Q = I$, $R=0.001I$ and $r(t)$ is a step function.
\begin{figure}[ht]
      \centering
     \includegraphics[scale = 0.55]{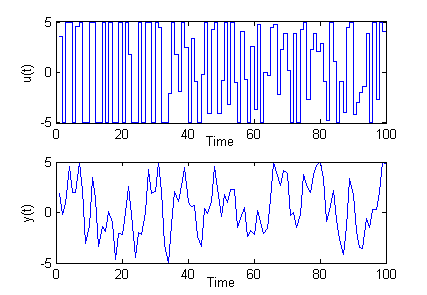}
      \caption{Input (top) and noiseless output (top), generated by the proposed method.} 
      \label{Figure:IO}
\end{figure}
The proposed algorithm has been applied to the problem and the resulting input and output signals are shown in Figure \ref{Figure:IO}. For the obtained input sequence, the constraint $I_F(\thetao) > \frac{\gamma \chi^2_\alpha(n_\theta)}{2}V^{''}_{app}(\thetao)$ is satisfied.
This can also be confirmed by checking the eigenvalues of the slack variable $S_{opt}$, which are all positive and the zero cost function.
%~~~~~~~~~~~~~~~~~~~~~~~~~~~~~~~~~~~~~~~~~~~~~~~~~~~~~~~~~~~~~~~~~~~~~~~~~~~~~~~~~
\section{Conclusion}
\label{sec:conclusion}
In this paper we introduced a new approach to generate input signals such that the estimated model based on the generated signal can guarantee a desired control performance. The method is based on satisfying a lower bound on the Fisher information matrix. The experimental cost is considered to be the minimum required time for satisfying the lower bound. One significant feature of the proposed approach is that the problem is formulated in time-domain and thus it is straightforward to handle constraint on the amplitude of the input and output signals. 

The problem is, however, highly non-convex. This is addressed through alternating optimization methods, where we are alternating between optimizing cost function for each variable while the others are fixed. We perform a time recursive algorithm and each optimization problem is solved in a receding horizon framework. As a result, the method can also be used with control. The algorithm terminate when the application requirement is satisfied. We can converge to a local minimum with this sort of optimization algorithms. However, numerical examples showed the method is consistent with previous results in the literature and general enough to be applied to any linear system structure.

Future research directions include extending the method to the closed loop system identification and integrate it to Model Predictive Control and we aim to design optimal input while at the same time we are concerning about the control performance. More extensions could be robust and adaptive approaches as expressed in Remark 1.

%~~~~~~~~~~~~~~~~~~~~~~~~~~~~~~~~~~~~~~~~~~~~~~~~~~~~~~~~~~~~~~~~~~~~~~~~~~~~~~~~~
%\begin{ack}
%The research leading to these results has received funding from the European Union’s Seventh Framework Programme (FP7/2007-2013) under grant agreement no 257059, the ’Autoprofit’ project (www.fp7-autoprofit.eu).
%\end{ack}

\bibliography{Bibliography/CDCref}

%\appendix
%\section{A summary of Latin grammar}    % Each appendix must have a short title.
%\section{Some Latin vocabulary}              % Sections and subsections are supported
                                                                         % in the appendices.
\end{document}